\documentclass[%
reprint,
superscriptaddress,
showpacs,
amsmath,amssymb,
prl,
]{revtex4-2}
\usepackage{graphicx}
\usepackage{dcolumn}
\usepackage{bm}
\usepackage{CJKutf8}
\usepackage{color}
\usepackage{amssymb}
\usepackage{amsfonts}
\usepackage{amsthm}
\usepackage{palatino}
\usepackage[colorlinks,urlcolor=blue,linkcolor=blue,citecolor=blue,anchorcolor=blue]{hyperref}
\makeatletter

\newcommand{\Rmnum}[1]{\expandafter\@slowromancap\romannumeral #1@}

\makeatother
\begin{document}
	
\begin{CJK*}{UTF8}{gbsn}
		
\preprint{APS/123-QED}
		
\title{Reconfigurable Topological Dissipative Light Bullets}
		
\affiliation{Ministry of Education Key Laboratory for Nonequilibrium Synthesis and Modulation of Condensed Matter, Shaanxi Province Key Laboratory of Quantum Information and Quantum Optoelectronic Devices, School of Physics, Xi'an Jiaotong University, Xi'an 710049, China}
\affiliation{Key Laboratory for Physical Electronics and Devices, Ministry of Education, School of Electronic Science and Engineering, Xi'an Jiaotong University, Xi'an 710049, China}
\affiliation{Institute of Spectroscopy, Russian Academy of Sciences, Troitsk, Moscow, 108840,	Russia}
\affiliation{Institut Universitari de Matem\`atica Pura i Aplicada, Universitat Polit\`ecnica de Val\`encia, Val\`encia, Spain}
\affiliation{The two authors contribute equally to this work.} 
		
\author{Qian Tang (唐茜)}
\affiliation{Ministry of Education Key Laboratory for Nonequilibrium Synthesis and Modulation of Condensed Matter, Shaanxi Province Key Laboratory of Quantum Information and Quantum Optoelectronic Devices, School of Physics, Xi'an Jiaotong University, Xi'an 710049, China}
\affiliation{The two authors contribute equally to this work.} 
		
\author{Yiqi Zhang (张贻齐)}
\email[Contact author: ]{zhangyiqi@xjtu.edu.cn}
\affiliation{Key Laboratory for Physical Electronics and Devices, Ministry of Education, School of Electronic Science and Engineering, Xi'an Jiaotong University, Xi'an 710049, China}
\affiliation{The two authors contribute equally to this work.} 
		
\author{Yaroslav V. Kartashov}
\affiliation{Institute of Spectroscopy, Russian Academy of Sciences, Troitsk, Moscow, 108840,	Russia}
		
\author{Carles Mili\'an}
\email[Contact author: ]{carmien@upvnet.upv.es}
\affiliation{Institut Universitari de Matem\`atica Pura i Aplicada, Universitat Polit\`ecnica de Val\`encia, Val\`encia, Spain}

\date{\today}
		
\begin{abstract}
\noindent
We discover a novel class of dissipative light bullets whose spatial profiles may be drastically reshaped along the bulk, edges and corners of the Su-Schrieffer-Heeger lattice owing to the dissipative and topological nature of the system. These light bullets appear due to resonances with different modes in lattice spectrum and may have very rich shapes that can be adiabatically controlled by varying the frequency of the external laser source. We report on robust stationary bullets and breathers which are understood from the corresponding bifurcation analysis. Our results provide a new route to realisation of reconfigurable and robust 3D light forms in topologically nontrivial systems.
%
\end{abstract}
		
\maketitle
		
\end{CJK*}
 
Topological insulator is a new phase of matter that only allows the surface of material to conduct with the bulk remaining insulating~\cite{hasan.rmp.82.3045.2010, qi.rmp.83.1057.2011}. The concept of topological insulators originates from solid state physics, but it has extended and received considerable development in photonics~\cite{lu.np.8.821.2014, ozawa.rmp.91.015006.2019, smirnova.apr.7.021306.2020, yan.aom.2001739.2021, segev.nano.10.425.2021, zhang.nature.618.687.2023, szameit.np.20.905.2024} and many other fields. In addition to photonic insulators belonging to rich classes with broken time-reversal~\cite{wang.nature.461.772.2009, rechtsman.nature.496.196.2013} or inversion~\cite{wu.nc.8.1304.2017, noh.prl.120.063902.2018} symmetry, there also exist higher-order topological insulators (HOTIs)~\cite{benalcazar.prb.96.245115.2017, xie.nrp.3.520.2021, lin.nrp.5.483.2023} supporting topological states with dimensionality at least by two lower than that of the system~\cite{noh.np.12.408.2018, hassan.np.13.697.2019, ni.nm.18.113.2019, xue.nm.18.108.2019, zhang.elight.3.5.2023}, whose topological properties can be described by the polarization index~\cite{xie.prl.122.233903.2019, chen.prl.122.233902.2019}. Photonic topological systems are particularly advantageous because they allow investigation of nontrivial interplay between topology and nonlinearity~\cite{mukherjee.science.368.856.2020, maczewsky.science.370.701.2020, mukherjee.prx.11.041057.2021, kirsch.np.17.995.2021, hu.light.10.164.2021, ren.light.12.194.2023, arkhipova.sb.68.2017.2023} and non-Hermitian effects~\cite{luo.prl.123.073601.2019, zhang.nc.12.5377.2021}. Thus, strong localization of corner states in HOTIs in comparison with states extended along the edge of the insulator~\cite{rechtsman.nature.496.196.2013, leykam.prl.117.143901.2016, klembt.nature.562.552.2018, noh.prl.120.063902.2018, lustig.nature.567.356.2019, zhong.ap.3.056001.2021} makes them advantageous for realization of the low-threshold corner lasers~\cite{zhang.light.9.109.2020, kim.nc.11.5758.2020, zhong.apl.6.040802.2021} and high-Q nanocavities~\cite{ota.optica.6.786.2019, xie.lpr.14.1900425.2020}.

Nonlinearity in topological systems serves as a powerful knob allowing to control evolution of topological states and leading to the formation of unique objects --- topological solitons~\cite{smirnova.apr.7.021306.2020, szameit.np.20.905.2024}. However, examples of nonlinear control of topological states, including in HOTIs~\cite{kirsch.np.17.995.2021, hu.light.10.164.2021}, were so far reported mostly in one- (1D) and two-dimensional (2D) settings, but they remain practically unexplored in 3D systems, except for recent theoretical prediction of 3D light bullets in conservative topological system~\cite{ivanov.pra.107.033514.2023}. Notice that observation of stable 3D solitons~\cite{silberberg.ol.15.1282.1990} remains one of the major open problems in photonics, since even the most fruitful previous approaches based on utilization of nontopological lattices~\cite{aceves.ol.19.329.1994, aceves.prl.75.73.1995, mihalache.pre.70.055603.2004, kevrekidis.prl.93.080403.2004, mihalache.prl.95.023902.2005, leblond.pre.76.026604.2007, malomed.job.7.r53.2005, grelu.np.6.84.2012, malomed.epjst.225.2507.2016, mihalache.rrp.69.403.2017, shtyrina.pra.97.013841.2018, kartashov.nrp.1.185.2019, rosanov.epjd.73.141.2019} allowed to observe light bullets only in transient form and in narrow energy range~\cite{minardi.prl.105.263901.2010, eilenberger.prx.3.041031.2013, renninger.nc.4.1719.2013} due to instabilities and influence of higher-order effects.

\begin{figure*}[ht]
\centering
\includegraphics[width=\textwidth]{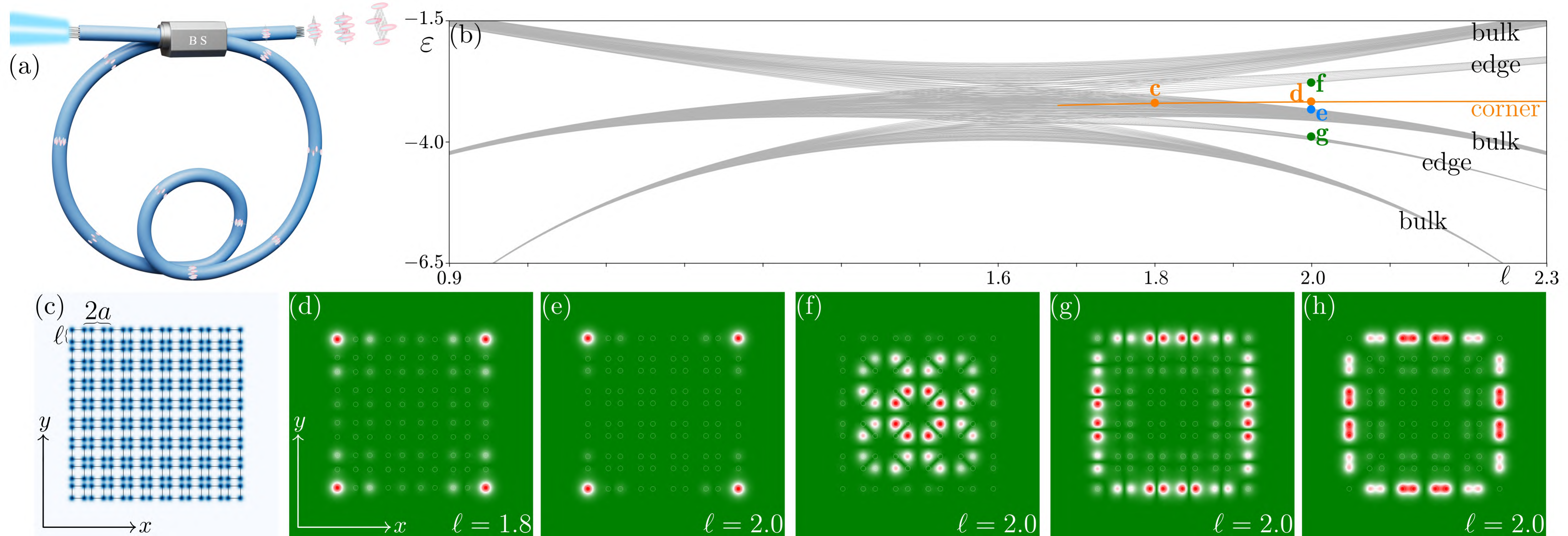}
\caption{(a) Sketch of the cavity with SSH lattice supporting the formation of bullets. BS represents the beam splitter. (b) Spectrum of the 2D SSH lattice as a function of $\ell$. 
(c) Profile of the 2D SSH lattice. (d-h) Spatial $|\psi|$ distributions for corner, bulk, and edge states corresponding to the dots in (b). Faint white circles mark the waveguide array. Panels in (c-h) are shown in ${-12\le x,y \le 12}$.}
\label{fig1}
\end{figure*}

Topological light bullets were never reported in dissipative driven systems. At the same time, such systems are known for their remarkable stability and unique opportunities for control of shapes and localization of topological excitations (see, e.g.~\cite{kartashov.optica.3.1228.2016, kartashov.prl.119.253904.2017, zhang.lpr.13.1900198.2019, zhang.ol.45.4710.2020, pernet.nphys.18.678.2022}). In this Letter, we report on robust corner, edge and bulk light bullets as well as the corresponding robust breathers in a driven dissipative Su-Schrieffer-Heeger (SSH) lattice system, the sketch of which is shown in Fig.~\ref{fig1}(a). We find the conditions under which all the above states coexist and are stable in a fixed lattice geometry, while reshaping one into another is easily achieved by tuning the cavity-laser detuning. The formation of light bullets in dissipative cavity systems has been previously reported in multi-mode single channel waveguides \cite{sun.prl.131.137201.2023, guo.nc.14.2029.2023, cao.light.12.260.2023} where the spatial reshaping of the bullet may occur only within a fixed core so that it only affects the out-coupling efficiency. In stark contrast with the above, we achieve a drastic spatial reshaping along the edges and corners of the lattice by exciting different kinds of topological bullets. Because each channel along the surface is individually accessible, this unprecedented bullet reshaping can be neatly detected at the output port achieving hence a truly spatially multiplexed output signal from our system. The stabilizing role of the lattice with small index contrast is central for our findings, since bullets in the damped driven Schr\"{o}dinger equation without a potential are all unstable \cite{milian.prl.123.133902.2019}.

The 3D light propagation in our system obeys the damped driven nonlinear Schr\"{o}dinger equation:
\begin{equation} \label{eq1}
i \partial_z \psi=-\frac{1}{2}\Delta \psi -\mathcal{R}(x, y) \psi-|\psi|^2 \psi  -i \gamma\psi - \varepsilon \psi + \mathcal{P},
\end{equation}
where $z$ is the propagation direction, ${\Delta\equiv \partial_x^2+\partial_y^2+\partial_t^2}$ is the 3D Laplacian, $t$ is the fast time~\cite{haelterman.oc.91.401.1992}, the function ${\mathcal{R}}(x,y)$ stands for the refractive index profile of the lattice, $\gamma$ is the loss parameter, $\varepsilon$ the cavity-laser detuning and ${\mathcal{P}}$ accounts for the cavity coupling with the external pump field. Here ${\mathcal{R}}(x,y)$ is composed of identical waveguides of width $\sigma$ placed in the nodes ${(x_{mn}, y_{mn})}$ of the 2D SSH grid ${{\mathcal{R}}(x,y) = p \sum_{mn} \exp\{- [(x - x_{mn})^2 + (y - y_{mn})^2]/\sigma^2 \}}$ characterised by a lattice constant ${a=1.6}$ [cf. Fig.~\ref{fig1}(c)] and a depth ${p = k^2 r_0^2 \delta n / n_0}$, where $\delta n$ is the refractive index contrast, $r_0$ is the characteristic transverse scale that is used to normalize $(x,y)$, ${k(\omega) = 2\pi n_0(\omega) /\lambda = n_0(\omega) \omega / c}$ is the wavenumber, $n_0$ is the ambient refractive index,
and $\lambda$ is the wavelength. We set ${\sigma=0.5}$ and ${p=10}$ that corresponds to typical parameters ${\delta n \sim 1.0\times 10^{-3}}$, ${n_0=1.45}$, ${r_0=20\,\mu \rm m}$, ${\lambda=1550\,\rm nm}$~\cite{minardi.prl.105.263901.2010}. The distance $z$ is normalized to $kr_0^2{\sim 2.35\,\rm mm}$, the normalized temporal coordinate ${t = (T - k r_0^2 z/v_g)/T_s}$ is defined in the co-moving frame with the group velocity $v_g$; ${T_s=[-\beta_2 k(\omega) r_0^2]^{1 / 2}}$ with ${\beta_2\equiv\partial^2 k/\partial \omega^2\approx -28\, \rm fs^2/mm}$ accounts for anomalous group velocity dispersion (GVD) leading to ${T_s = 8.1\,\rm fs}$. Below we set losses to ${\gamma=0.01}$ and pump amplitude to ${\mathcal{P}=0.004}$~\footnote{The particular choice of parameters $\{p,a,\sigma,\gamma,\mathcal{P}\}$ is critical in order to correspond to a realistic realisation and, most importantly, to achieve that bullets of the different types, namely bulk bullets and topological bullets (corner and edge) coexist and are stable over a generous region of the parameter space for a fixed lattice geometry. The latter constitutes the central result of this work.}. The topological properties of the 2D SSH lattice are controlled by the variable inter-cell spacing $\ell$ [cf.~Fig.~\ref{fig1}(c)] which strongly impacts the lattice spectrum [cf.~Fig.~\ref{fig1}(b)] as well as the linear modes [cf.~Figs.~\ref{fig1}(d)-\ref{fig1}(h)]~\footnote{The results can be obtained by numerically solving the spatial eigenvalue problem associated to Eq.~(\ref{eq1}) in the linear and conservative limit: ${[\frac{1}{2}\Delta+\mathcal{R}(x,y)]\psi_n=-\varepsilon\psi_n}$, using the plane-wave expansion method}.

 \begin{figure*}[ht]
\centering
\includegraphics[width=\textwidth]{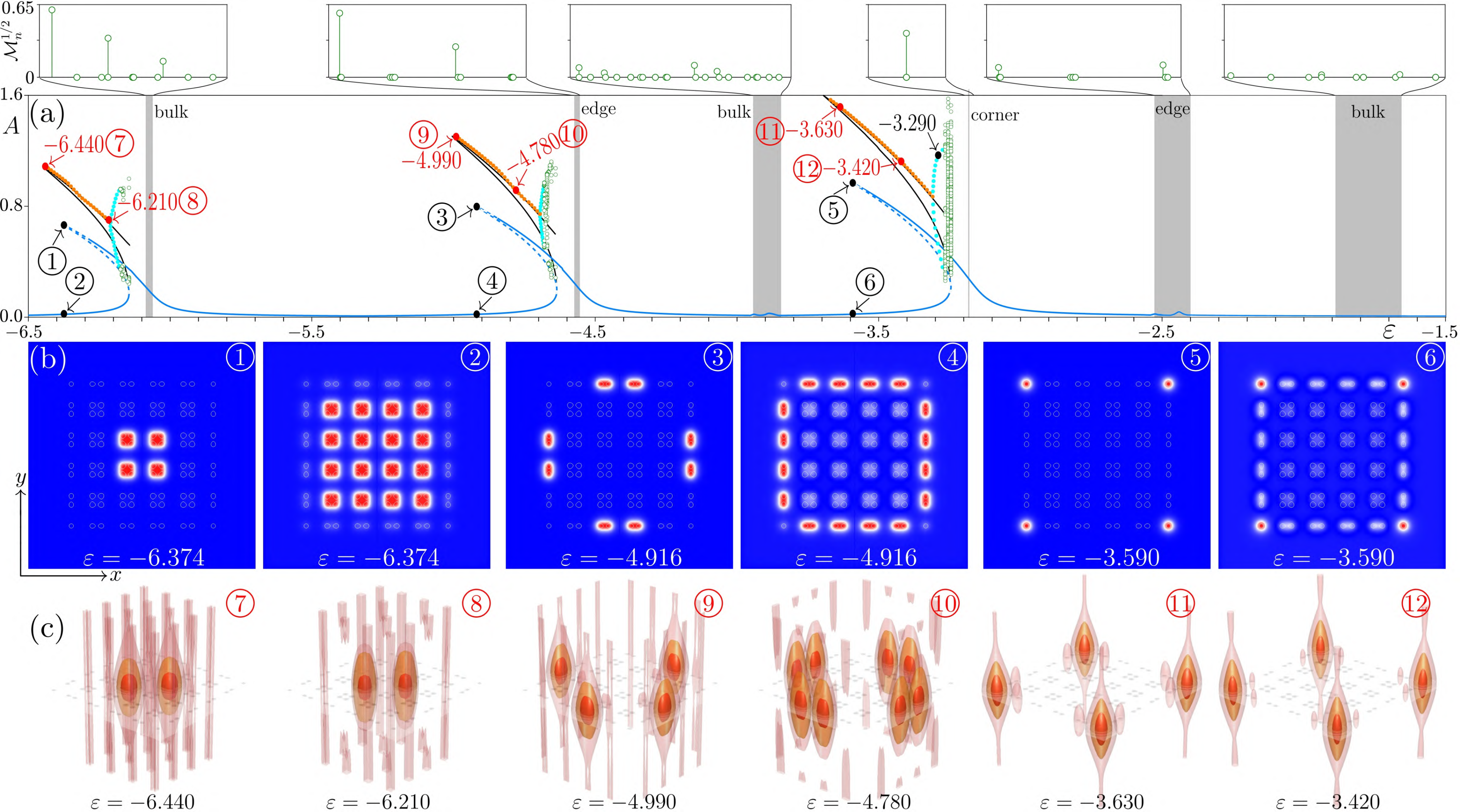}
\caption{(a) Peak amplitude $A$ of the bistable nonlinear states as a function of $\varepsilon$ at ${\ell=2.2}$ 
(blue curve). Unstable states are shown by dashed blue curves. 
Orange dots show $A$ of stable bullets, the cyan dots show bullet breathers (the upper and lower ones are the maximum and minimum peak amplitudes, respectively), while the hollow green dots show the chaotic beams. 
Black curves represent $A$ of the bullets obtained with iteration method. (b) $|\psi|$ distributions of states corresponding to the black dots with circled numbers in (a). (c) Isosurface plots of the bullets for ${-9.5\le x,y,t\le9.5}$ corresponding to the red dots with circled numbers in (a). Isosurface levels are $0.02$ (pink), $0.1$ (orange), and $0.5$ (red).}
\label{fig2}
\end{figure*}
	
\begin{figure*}[ht]
\centering
\includegraphics[width=\textwidth]{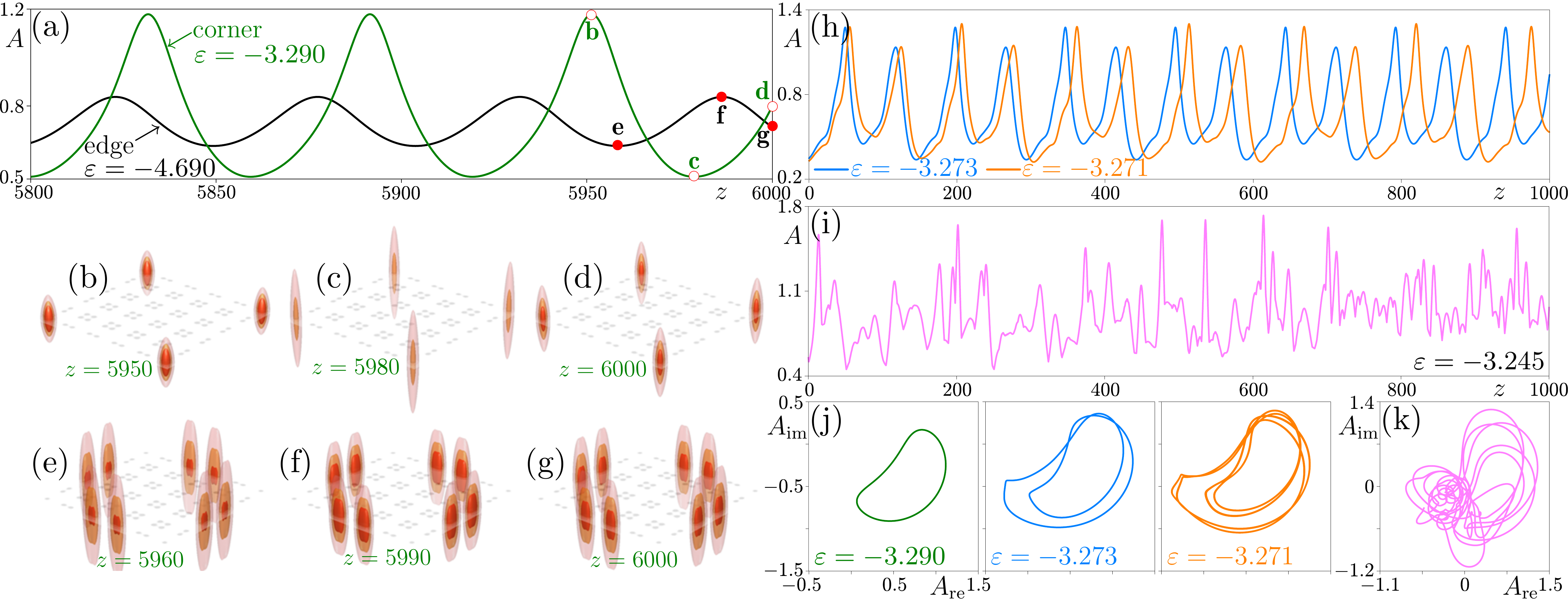}
\caption{(a) Peak amplitude $A$ of the bullet breathers 
during a propagation initialised at ${z=0}$. (b-d) Isosurface plots of the corner bullet breathers corresponding to the hollow red dots on the green curve in (a). (e-g) Edge bullet breathers corresponding to the solid red dots on the black curve in (a). Levels for all isosurface plots are $0.2$ (pink), $0.4$ (orange), and $0.6$ (red). (h) Peak amplitudes of the corner bullet breathers with ${\varepsilon=-3.273}$ (blue) and ${-3.271}$ (orange). (i) Peak amplitude of the 3D chaotic beam with ${\varepsilon=-3.245}$. (j) Orbits formed by the real $A_{\rm re}$ and imaginary ${A_{\rm im}}$ parts of $A$ in (a,\,h). (k) Orbit of $A$ in (i).
}
\label{fig3}
\end{figure*}

	
	
In our optical cavity geometry, the parameter $\varepsilon$ plays the role of the cavity-laser detuning so that prominent transmission resonances featuring bistability [Fig.~\ref{fig2}(a)] may appear around the frequencies $\varepsilon$ of the driving field matching the eigenfrequencies of certain cavity modes. Figure~\ref{fig2}(a) shows three such resonances (blue curves) for the peak amplitude of the intracavity field ${A=\max\{|\psi|\}}$ versus detuning, tilted due to the nonlinearity, bifurcating from linear bulk, edge, and corner modes of the SSH lattice in topological regime ${\ell=2.2}$. The existence of these resonances is subject to an efficient overlap between the pump and the intracavity modes given by ${{\mathcal{M}}_n\equiv\int\hspace{-0.5em}\int {\mathcal{P}}\psi^*_n(x,y)\mathrm{d}x\mathrm{d}y}$ [see Fig.~\ref{fig2}(a), top insets]. Modes with poor overlap do not rise resonances efficiently, as clearly seen for the modal bands around ${\varepsilon\approx-3.9}$, $-2.5$, and $-1.8$ in Fig.~\ref{fig2}(a). Our particular choice of a flat pump field brings a remarkable heterogeneous scenario where several resonances form and are neatly separated in $\varepsilon$. Each of these resonances consists of 2D nonlinear monochromatic states with clear bulk, edge and corner origins; see examples in Fig.~\ref{fig2}(b). The monochromatic fields are found numerically with Newton-Raphson method by seeking solutions of Eq.~(\ref{eq1}) of the form ${\partial_z\psi=\partial_t^2\psi=0}$. The nonlinear monochromatic modes are robust against \textit{transverse} instabilities along the solid blue curves in Fig.~\ref{fig2}(a), while they are all unstable to \textit{temporal} modulations at ${A\gtrsim0.2}$. 
	
	
While in 1D and 2D passive cavities the combination of nonlinear resonances (or bistability) with anomalous GVD is closely linked to the formation of stable solitonic states (see, e.g.~\cite{barashenkov.pre.54.5707.1996,firth.josab.19.747.2002,milian.prl.121.103903.2018}), this cannot be guaranteed in 3D counterparts due to dramatic decrease of stability region with the increase of dimensionality~\cite{barashenkov.pre.54.5707.1996,firth.josab.19.747.2002} to the point that Eq.~(\ref{eq1}) with ${p\to 0}$ does not allow stable 3D solitons~\cite{milian.prl.121.103903.2018}. Surprisingly, however, we find very robust 3D light bullets associated to each of the above resonances and in all three cases over a generous region of the parameter space. Bullet branches of bulk, edge and corner natures are shown in Fig.~\ref{fig2}(a), where the latter two are of pure topological origin. Light bullets are found both with direct propagation of Eq.~(\ref{eq1}) (orange dots) as well as with a 3D Newton-Raphson solving Eq.~(\ref{eq1}) by imposing ${\partial_t\psi=0}$ (black lines). Examples of stable bullets are shown in Fig.~\ref{fig2}(c). The three bullet families are robust and coexist within the same cavity geometry so that transition amongst them is possible simply by adjusting the detuning $\varepsilon$. The existence of these novel rich and re-configurable bullets constitutes our central result.
	
\begin{figure*}[htbp]
\centering
\includegraphics[width=\textwidth]{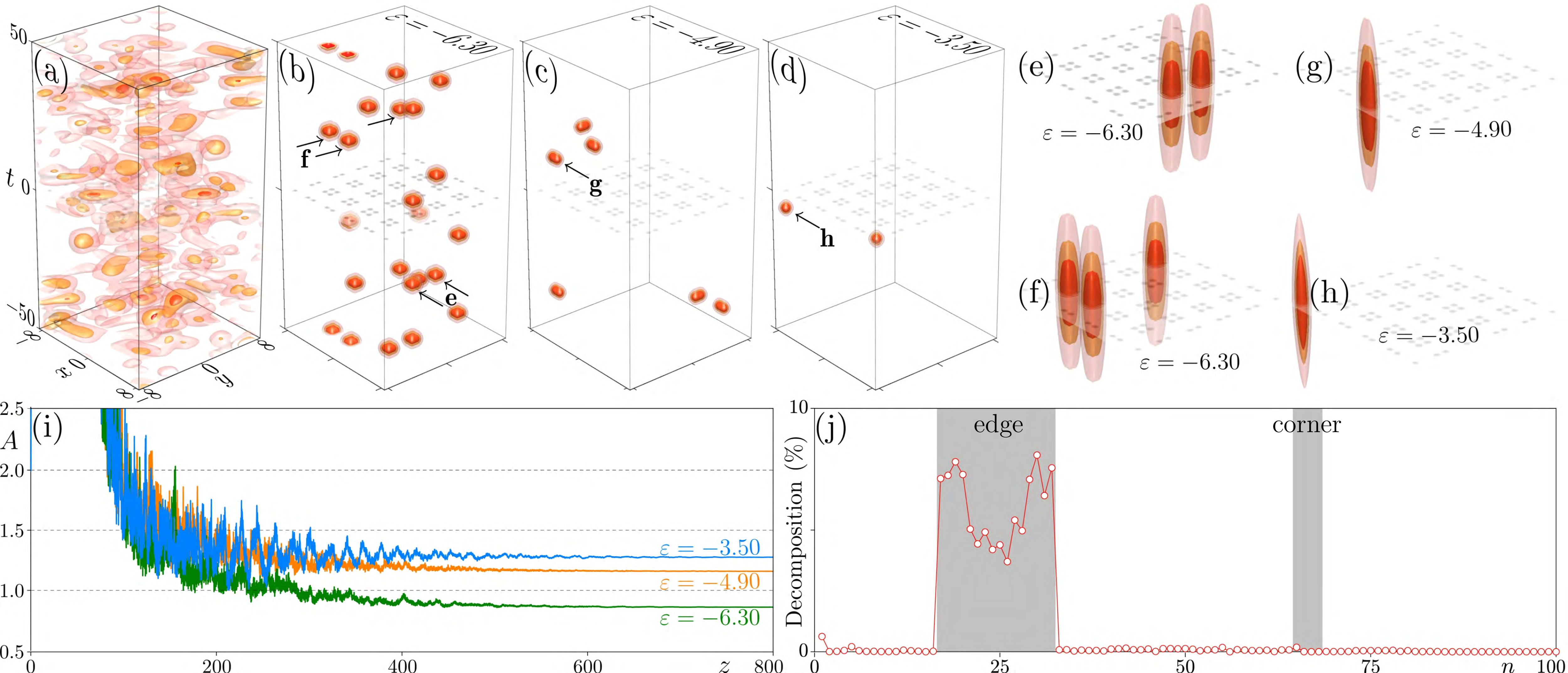}
\caption{(a) Isosurface of the noisy input. (b) Excitation of the bulk bullets (with maxima at different times $t$) with ${\varepsilon=-6.30}$ at ${z=800}$.
Excitation of the edge bullets at ${\varepsilon=-4.90}$ (c) and corner bullets at ${\varepsilon=-3.50}$ (d). Isosurface levels are $1.0$ (pink), $1.4$ (orange), and $1.8$ (red) in (a), and $0.2$ (pink), $0.4$ (orange), and $0.6$ (red) in (b,c,d). (e,f) Selected bulk bullets indicated by arrows in (b), (g) edge bullet from (c), and (h) corner bullet from (d). (i) Peak amplitude of the noisy input during propagation for different $\varepsilon$. (j) Decomposition of the edge bullet in (g) in the ($x,y$) cross section on all linear modes.}
\label{fig4}
\end{figure*}
	
Stability of light bullets is found only for the top $A(\varepsilon)$ branches in Fig.~\ref{fig2}(a) and it is checked numerically by long propagation runs of Eq.~(\ref{eq1}) up to ${z\gtrsim 6000}$ seeded with the corresponding bullet plus a small-scale random perturbation of ${\sim5\%}$ in amplitude. When bullets are stable, these simulations display $A(z)$ traces converging to a constant value [as in Fig.~\ref{fig4}(i)], represented in the orange dots in Fig.~\ref{fig2}(a) for bulk, edge and corner bullets. Upon the increase of $\varepsilon$, all bullet families evolve into stable breathers that persist for arbitrarily long distances displaying oscillatory $A(z)$ traces as in Fig.~\ref{fig3}(a). The maximum and minimum amplitudes are marked by the pair of cyan dots in Fig.~\ref{fig2}(a) emerging from the stable (orange) bullets [for even larger $\varepsilon$ values chaotic dynamics is encountered, shown by the hollow green dots]. The green trace in Fig.~\ref{fig3}(a) shows a perfectly periodic evolution of the amplitude of corner bullet breather (at ${\varepsilon=-3.29}$), depicted in Figs.~\ref{fig3}(b)-\ref{fig3}(d) for different $z$. In this case, the four spots on the corners oscillate in the perfectly synchronised fashion. Some breathers may display seemingly irregular amplitude oscillations like those shown in Fig.~\ref{fig3}(h) by the blue (${\varepsilon=-3.273}$) and orange (${\varepsilon=-3.271}$) curves. This peculiar behaviour arises because 2 pairs of spots in the diagonal direction do not oscillate synchronously. 
We note the remarkable robustness of these breathers despite strong oscillations along propagation. The example of bullet with ${\varepsilon=-3.245}$ demonstrating chaotic amplitude oscillations is shown in Fig.~\ref{fig3}(i).
The periodicity of the bullet breathers is well recognized from the trajectories of the real and imaginary parts of the peak amplitude during propagation [Fig.~\ref{fig3}(j)], showing one loop [corresponding to state in Fig.~\ref{fig3}(a)] or two and four loops [corresponding to Fig.~\ref{fig3}(h)]. Figure~\ref{fig3}(k) shows chaotic trajectory corresponding to Fig.~\ref{fig3}(i). The examples of edge bullet breather at ${\varepsilon=-4.69}$ is shown by the black curve in Fig.~\ref{fig3}(a) and in Figs.~\ref{fig3}(e)-\ref{fig3}(g).

	
It is important to note that all these light bullets are easily excited from random noise~\footnote{The noisy input is created in the way: $\psi(x,y,t;z=0) = 2 \mathcal{F}_{\mathcal N}^{-1} \{ \delta(k_x,k_y,k_t) \exp[- (k_x^2 + k_y^2 + k_t^2)/\sigma_k^2] \} $ where $\delta$ is a 3D small perturbation which is composed of random numbers, $k_{x,y,t}$ are the coordinates in the inverted space, $\sigma_k$ is the width, and $\mathcal{F}_\mathcal{N}^{-1}$ represents the inverse Fourier transform operator with the peak amplitude of the result after transform normalized.} example of which at $z=0$ is shown in Fig.~\ref{fig4}(a). Bulk, edge, and corner bullets excited in Fig.~\ref{fig4}(b)-\ref{fig4}(d), respectively, arise due to the proper choice of $\varepsilon$~\footnote{As anticipated by our results in Fig.~\ref{fig2}, the type of light bullet that forms is determined by the value of the cavity-laser detuning $\varepsilon$. Note that $\varepsilon$ plays the role of a propagation constant in non-driven systems and as such can be precisely chosen, as is the case here.}. Notice that in dynamically excited corner bullets not all 4 corners need to be occupied [Fig.~\ref{fig4}(d)]. Similar behaviour is observed for edge [cf.~Fig.~\ref{fig4}(c)] and bulk [cf.~Fig.~\ref{fig4}(b)] bullets. In Figs.~\ref{fig4}(e)-\ref{fig4}(h), we display bulk, edge and corner bullets chosen from those in Figs.~\ref{fig4}(b)-\ref{fig4}(d), as indicated by arrows~\footnote{The existence of bullet solutions of this types are checked with Newton-Raphson method, which shows that they all have very similar amplitude branches that are almost indistinguishable from those shown in Fig.~\ref{fig2}. }. Upon dynamical excitation from noise, (independent) bullets of the same type (defined by the value of $\varepsilon$) typically appear in different locations along $t$ axis [see example in Figs.~\ref{fig4}(e) and \ref{fig4}(f), where we show two zooms from the same multi-bullet bulk state around two different locations in $t$]. In Figs.~\ref{fig4}(g) and \ref{fig4}(h), we show the excited edge and corner bullets with only one spot, by zooming over a particular $t$ range. Curves in Fig.~\ref{fig4}(i) show $A(z)$ dependencies for bullet excitation process from the noisy inputs. All types of stable bullets form rather quickly, they reach final form around ${z\sim800}$ and persist over arbitrarily long propagation distances. We stress that edge bullets are particularly spectrally rich as seen from their decomposition over linear modes from SSH lattice spectrum in Fig.~\ref{fig4}(j)~\footnote{Such decomposition is obtained by projecting the bullet's cross section, $\psi_B(x,y,t_0)$ [across the $(x,y)$ plane], at the ${t=t_0}$ for which the $\max{|\psi_B(x,y,t)|}$ is found; see, e.g., Ref.~\cite{wang.prl.121.194301.2018}.}. Similarly decomposition of corner and bulk bullets reveals dominating contributions from corner and bulk states, respectively.
	
To conclude, we have reported on robust topological corner, edge, and bulk bullets coexisting for a given dissipative topological SSH lattice geometry with focusing nonlinearity and anomalous GVD. Remarkably robust dynamical bullet breathers are encountered too. Importantly, all these bullets can be excited from random noise by properly choosing the cavity-laser detuning, which is an inherent advantage of our dissipative system. These results illustrate that topologically nontrivial structures with rich modal spectrum offer powerful platform for generation of higher-dimensional self-sustained states with controllable spatial structure.
	
\textit{Acknowledgments}---This work was supported by 
the Natural Science Basic Research Program of Shaanxi Province (2024JC-JCQN-06),
the National Natural Science Foundation of China (12474337, 12074308), 
FFUU-2021-0003 of the Institute of Spectroscopy of the Russian Academy of Sciences, 
the Russian Science Foundation (21-12-00096), and
and the Fundamental Research Funds for the Central Universities (xzy012024135). C.M. acknowledges support from the Spanish government via grant PID2021-124618NB-C21, funded by MCIN/AEI/10.13039/501100011033 and ‘ERDF: a way of making Europe’ of the European Union and from Generalitat Valenciana PROMETEO/2021/082.

\bibliography{my_library}
	
\end{document}